\documentclass[trans]{IEEEtran}
\usepackage[T1]{fontenc}
\usepackage{hyperref}
\usepackage{lettrine}
\usepackage{epsfig}
\usepackage{float}
\usepackage{footnote}
\usepackage{cite}
\usepackage{float}
\usepackage{amssymb}
\usepackage{amsthm}
\usepackage{amsmath,mathabx}
\usepackage{color}
\usepackage{epstopdf}
\usepackage{tabularx,tabulary}
\usepackage[font=footnotesize]{caption}
\usepackage{graphicx}
\usepackage{caption}
\usepackage{subcaption}
\usepackage{fontawesome}
\usepackage{cuted}
\usepackage{dblfloatfix}    
\usepackage{bbm}
\usepackage{algorithm,algorithmic}
\usepackage{multirow}
\usepackage{wrapfig}

\renewenvironment{IEEEbiography}[1]
  {\IEEEbiographynophoto{#1}}
  {\endIEEEbiographynophoto}
\usepackage[table,xcdraw]{xcolor}

\begin{document}
\title{Wireless Data Center Networks: \\ Advances, Challenges, and Opportunities}
\author{
Abdulkadir~Celik,~\IEEEmembership{Member,~IEEE,} Basem~Shihadah,~\IEEEmembership{Senior Member,~IEEE,} and~Mohamed-Slim~Alouini,~\IEEEmembership{Fellow,~IEEE}
\thanks{The authors are with Computer, Electrical, and Mathematical Sciences and Engineering Division at King Abdullah University of Science and Technology (KAUST), Thuwal, KSA.}
}


\maketitle
\begin{abstract}
Data center networks (DCNs) are essential infrastructures to embrace the era of highly diversified massive amount of data generated by emerging technological applications. In order to store and process such a data deluge, today's DCNs have to deploy enormous length of wires to interconnect a plethora of servers and switches. Unfortunately, wired DCNs with uniform and inflexible link capacities expose several drawbacks such as high cabling cost and complexity, low space utilization, and lack of bandwidth efficiency. Wireless DCNs (WDCNs) have emerged as a promising solution to reduce the time, effort, and cost spent on deploying and maintaining the wires. Thanks to its reconfigurability and flexibility, WDCNs can deliver higher throughputs by efficiently utilizing the bandwidth and mitigate the chronic DCN problems of oversubscription and hotspots. Moreover, wireless links enhance the fault-tolerance and energy efficiency by eliminating the need for error-prone power-hungry switches. Accordingly, this paper first compares virtues and drawbacks of millimeter wave (mmWave), terahertz (THz), and optical wireless communications as potential candidates. Thereafter, an in-depth discussion on advances and challenges in WDCNs is provided including physical and virtual topology design, quality of service (QoS) provisioning, flow classification, data grooming, and load balancing. Finally, exciting research opportunities are presented to promote the prospects of WDCNs.    
\end{abstract}

\maketitle

\section{Introduction}
\label{sec:intro}
\lettrine{D}{ata Centers} (DCs) are essential infrastructures to process and store massive amounts of data generated by emerging technological trends such as fifth generation (5G) networks, Internet of things,  big data, cloud services, artificial intelligence, etc. The diverse quality of service (QoS) demands of these services (i.e., storage capacity, processing power, bandwidth, latency, etc.) can be fulfilled, either by centralized mega DCs scaling up to hundreds of servers, or by a network of distributed micro DCs. In this respect, there is a dire need for novel networking solutions, architectural approaches, and communication technologies to provide high-speed low-latency interconnections within and between DCs. Those are also referred to as intra-DC and inter-DC networking, respectively. Therefore, a scalable, resilient, and sustainable DCN design is of the utmost importance to utilize network resources efficiently, minimize capital and operational expenditures, and adapt network topology to dynamically changing traffic patterns.

Among many other architectures, technology giants (e.g., Google, Facebook, Amazon, etc.) construct DCNs in a hierarchical tree topology where servers are arranged in racks. While intra-rack communication is realized by top-of-rack edge switches (ESs) in the lower layer, inter-rack communication is established by linking ESs with aggregate switches (ASs) in the middle layer and then connecting ASs with the core switches (CSs) in the top layer. In traditional DCNs, all these connections are established via uniform and fixed capacity cables (e.g., coaxial, twisted-pair, fiber, etc.). However, this hierarchical topology yields a multi-root tree where more powerful links and switches are required in the branches nearer the top layer, which makes CSs the DCN bottleneck in heavy traffic conditions. Therefore, inter-rack communication observes a throughput much lower than the actual available bandwidth \cite{skandula2009flyways}, a.k.a. \textit{oversubscription}. Moreover,  measurements and analysis of real-life DCN traffic characteristics reveal that some applications generate unpredictable traffic patterns and asymmetrical traffic distributions \cite{skandula}. This unbalanced traffic is mainly caused by \textit{hotspots} that contain common data required by many ongoing jobs in several DCN entities. The DCN traces also show that only 60\% of the edge and core links are active at a time and utilization of 95$^{th}$ percentile of aggregation links is below 10\% \cite{Benson2010Understanding}.

It is challenging for wired DCNs to conform with this unpredictable and unbalanced traffic because of the fixed hierarchical topology and the inflexible links with uniform capacity. Existing efforts deal with these phenomena either by extending the aforementioned switch-centric hierarchical tree topology (e.g., Fat-Tree, VL2, PortLand, etc.) or by designing new server-based topologies (e.g., BCube, DCell, FiConn,  Jellyfish, etc.). Nonetheless, these approaches are still inadequate to prevent performance degradation caused by stochastic and unbalanced traffic conditions. Although additional wired on-demand links are a quick solution to increase the capacity, it is hard to predict location of oversubscription and hotspots. On the other hand, adding extra cables for each server is not practical from the technical and economical points of view. As the size of DCN scales up, the cabling cost and complexity increase along with indirect consequences such as: incompetent cooling because of the heat dissipation; inefficient space utilization due to the thick cable bundles; and operational costs of cabling management, maintenance, and modification. 

As a remedy, WDCNs have recently attracted significant attention to augment the aforementioned limitations of the wired DCs. Motivated by the fact that more than 95\% of the entire DCN traffic is carried out by the top 10\% largest flows, Skandula \textit{et. al.} were first to propose mitigating hotspots by establishing wireless on-demand links in DCNs oversubscribed with such large flows \cite{skandula2009flyways}. WDCNs can support reconfigurability for adapting the DCN to dynamically changing traffic patterns and loads; deliver higher throughputs via flexible links; reduce the capital investment and operational costs; offer convenient deployment, management, and maintenance; and enhance power efficiency by eliminating the need for switches. Nevertheless, realizing WDCNs poses formidable challenges such as designing obstruction-free physical topology, adjusting the virtual topology to dynamically changing traffic conditions, optimizing the link attributes to meet the QoS demands of different flow classes, tackling the power control and interference management, etc. 

In this regard, this paper first surveys and compares the virtues and drawbacks of potential technologies such as mmWave, THz, and optical wireless communications (OWCs). Then, present recent advances and provide important insights into open research problems in the literature. Finally, we relish exciting prospects of WDCNs and conclude the paper with a few remarks.

\section{Overview  of High-Speed Wireless Technologies}
\label{sec:overview}
Wireless technologies provide unique DCN attributes which is summarized as follows:

\begin{itemize}
\item 
Wiring a huge number of servers is a quite demanding engineering task which exacerbates as the DCN size scales up. Since wire deployment is usually an error-prone task, necessary maintenance and modifications also consumes considerable time and efforts. On the other hand, convenient plug-and-play wireless modules can alleviate wiring cost and complexity to a great extent. Removing wires also increases the cooling efficiency and space utilization.

\item 
Wireless techniques yield a more energy and bandwidth efficient DCN by eliminating the need for switches which are typically power-hungry, error-prone, and bandwidth limited. 
\item 
WDCNs can handle oversubscription and hotspots thanks to its flexibility and reconfigurability. Given an effective hardware topology, adapting the virtual topology as per the QoS demands and traffic loads also provides a higher throughput and efficient bandwidth utilization.
\end{itemize}
Nonetheless, there exist formidable challenges in design and provisioning of WDCNs, which varies with distinct characteristics of the underlying communication technology. In the remainder, we present the virtues and drawbacks of potential wireless techniques along with a detailed comparison. 

\subsection{mmWave Communications}
The current research efforts on the extremely high frequency (EHF) band (30-300 GHz) mostly focus on the 60 GHz band (57-64 GHz) and the E-band (71–76 and 81–86 GHz). With a proper frequency block granularity and efficient resource allocation, the available EHF spectrum can be sufficient to provide the mega-DCNs with the required level of scalability. However, there are fundamental discrepancies between mmWave and microwave bands in terms of propagation loss, penetration, and directivity. Despite of the high bandwidth, mmWave suffers from short communication ranges due to high propagation loss resulting from strong interactions with atmospheric constituents. 

Thanks to small wavelengths, mmWave links are directional by nature and suitable to implement electronically steerable antenna arrays with significantly high gains even on a small form factor transceiver. Therefore, beamforming is regarded as a key technique to extend communication range, suppress the interference, multiplex different users on the same beam, and support multi-beam directional operation for a higher spatial diversity. It is worth mentioning that lack of a proper alignment between transceivers causes a \textit{deafness} problem due to the high directivity. While deafness is critical for establishing links especially in mobile environments, it is also beneficial to diminish the interference as the receiver is tuned to a specific spatial channel. Nonetheless, high penetration loss of mmWave channels due to the scatterers yields a blockage problem, and thus line-of-sight (LoS) links are required. 

\subsection{TeraHertz Communications}

Being located between the EHF and infrared band (i.e., 0.1- 10 THz), the THz band offers high-speed communication, varying between tens of Gbps and several Tbps, depending on the link distance. Although the EHF and THz bands have common propagation characteristics, the THz communication has its unique advantages and implementation challenges. For instance, the propagation loss of the THz band is more severe and frequency selective than that of mmWave, which is mainly characterized by spreading and absorption effects. For frequencies below 1 THz, absorption effects are almost negligible for distances below 1 m whereas several transmission windows appear for distances above 1 m such as (0.38-0.44, 0.45-0.52, 0.62-0.72, 0.77-0.92) THz. Thanks to small footprints (i.e., the wavelengths ranging between 1 $\rm{mm}$ and 100 $\rm{\mu m}$), the severe propagation losses can be compensated by interleaving a large number of antennas of different bands in small form factor transceivers. By employing effective steering and beamforming techniques, it is possible to obtain collimated THz beams for achieving a substantial data rate in short to medium ranges and significantly reducing the interference as a result of the high directionality.  

As a matter of the course, there exist daunting challenges in the design and development of ultra-high-speed transceivers and ultra-broadband antennas. Silicon Germanium and Gallium Nitride technologies are already studied to achieve high power and low noise figure THz transceivers with desirable sensitivity levels. Additionally, Graphene is known as "the wonder material" with its superior electro-optical properties and considered to be one of the most potential alternatives for future THz modules. Nevertheless, existing research and development activities are still in its early stage and more efforts are needed to see ready to install commercial products.

\subsection{Optical Wireless Communications}
OWC refers to transmission in the unguided medium by using visible, infrared, and ultraviolet light-beams as a signal carrier. In particular, OWC operates in the near-infrared band is referred to as free-space optical (FSO) communications which can provide very high-speed links thanks to the plentiful bandwidth availability. Even though ongoing research efforts focus on tackling the outdoor FSO channel impediments (e.g., atmospheric turbulence, scintillation, misalignment and pointing errors, etc.), acclimatized indoor environment and reasonable range requirements of DCNs provide a much more tolerant playground for the use of FSO technology. Note that this friendly environment also obviates the need for bulky and costly transceivers designed for outdoor. 

\begin{table}
\caption{\footnotesize Comparison of Wireless Technologies}
\label{tab:comp}
\centering
\resizebox{0.5\textwidth}{!}{
\begin{tabular}{cccc}
\hline
\rowcolor[HTML]{34CDF9} 
\multicolumn{1}{|c|}{\cellcolor[HTML]{34CDF9}\bf Attributes}   & \multicolumn{1}{c|}{\bf mmWave}& \multicolumn{1}{c|}{\bf TeraHertz ( < 1 THz)} & \multicolumn{1}{c|}{\bf Indoor FSO} \\ \hline
&          &                       & \vspace{-7pt} \\ \hline
\rowcolor[HTML]{EFEFEF} 
\multicolumn{1}{|c|}{\cellcolor[HTML]{EFEFEF}\bf Bandwidth}   & \multicolumn{1}{c|}{$\approx$ 20 GHz} & \multicolumn{1}{c|}{$\approx$ 1 THz} & \multicolumn{1}{c|}{$\gg$ 100 THz }  \\ \hline
\rowcolor[HTML]{ECF4FF} 
\multicolumn{1}{|c|}{\cellcolor[HTML]{ECF4FF}\bf Regulation}   & \multicolumn{1}{c|}{No} & \multicolumn{1}{c|}{No} & \multicolumn{1}{c|}{No}  \\ \hline
\rowcolor[HTML]{EFEFEF} 
\multicolumn{1}{|c|}{\cellcolor[HTML]{EFEFEF}\bf Path Loss}    & \multicolumn{1}{c|}{\begin{tabular}[c]{@{}c@{}}High\\ $\approx$ 70 (90) dB for 1 (10) m
\end{tabular}} 
& \multicolumn{1}{c|}{\begin{tabular}[c]{@{}c@{}}Very High\\ $\approx$ 90 (110) dB for 1 (10) m\end{tabular}} & \multicolumn{1}{c|}{\begin{tabular}[c]{@{}c@{}}Medium\\ $\approx$ 0 dB for 10 m\end{tabular}} \\ \hline
\rowcolor[HTML]{ECF4FF} 
\multicolumn{1}{|c|}{\cellcolor[HTML]{ECF4FF}\bf Penetration Loss}   & \multicolumn{1}{c|}{High} & \multicolumn{1}{c|}{Very High} & \multicolumn{1}{c|}{Extremely High}  \\ \hline
\rowcolor[HTML]{EFEFEF} 
\multicolumn{1}{|c|}{\cellcolor[HTML]{EFEFEF}\bf Range}        
& \multicolumn{1}{c|}{Short-Medium( $\leq 30$ m)}      & \multicolumn{1}{c|}{Short ( $\leq 10$ m)}   & \multicolumn{1}{c|}{Medium-Long ( $\leq 1$ km)}     \\ \hline
\rowcolor[HTML]{ECF4FF} 
\multicolumn{1}{|c|}{\cellcolor[HTML]{ECF4FF}\bf Interference} & \multicolumn{1}{c|}{Very Limited} & \multicolumn{1}{c|}{Very Limited} & \multicolumn{1}{c|}{None} 
\\ \hline
\rowcolor[HTML]{EFEFEF} 
\multicolumn{1}{|c|}{\cellcolor[HTML]{EFEFEF}\bf Noise Source} & \multicolumn{1}{c|}{Thermal Noise} & \multicolumn{1}{c|}{Absorption \& Thermal Noise} & \multicolumn{1}{c|}{Ambient Light}                                                 
\\ \hline
\rowcolor[HTML]{ECF4FF} 
\multicolumn{1}{|c|}{\cellcolor[HTML]{ECF4FF}\bf Cost} & \multicolumn{1}{c|}{Low} & \multicolumn{1}{c|}{Low} & \multicolumn{1}{c|}{Low}\\ \hline
\rowcolor[HTML]{EFEFEF} 
\multicolumn{1}{|c|}{\cellcolor[HTML]{EFEFEF}\bf TRx Size}  & \multicolumn{1}{c|}{Small} & \multicolumn{1}{c|}{Very Small} & \multicolumn{1}{c|}{Small}
\\ \hline
\rowcolor[HTML]{ECF4FF} 
\multicolumn{1}{|c|}{\cellcolor[HTML]{ECF4FF}\bf Availability} & \multicolumn{1}{c|}{Available} & \multicolumn{1}{c|}{Emerging} & \multicolumn{1}{c|}{Available} 
 \\ \hline
\rowcolor[HTML]{EFEFEF} 
\multicolumn{1}{|c|}{\cellcolor[HTML]{EFEFEF}\bf Security} & \multicolumn{1}{c|}{High} & \multicolumn{1}{c|}{High} & \multicolumn{1}{c|}{High} 
 \\ \hline
\end{tabular}
}
\end{table}

Light emitting diodes (LEDs) and laser diodes (LDs) are two prominent OWC transmitters commonly used in practice. Since the LEDs are cheaper and more reliable light sources, visible light communication (VLC) can be used to cover a short-range and wide-beam coverage area with relatively low data rates. Therefore, multiple access schemes are necessary for a VLC receiver to distinguish signals coming from multiple LED sources. On the contrary, LDs generate razor-sharp light beams which can reach long distances at very high data rates. Even if LDs provide interference-free communication, precise alignment of transceivers is a must to establish LoS links. Furthermore, when it is combined with wavelength division multiplexing (WDM), FSO can provide very high data rates and fanout required by DCNs \cite{CelikSPIE}.

\subsection{Comparison Between Wireless Technologies}
Table \ref{tab:comp} compares three potential wireless technologies, none of which is regulated by governmental agencies for spectrum use. Given available bandwidths, it is possible to achieve Gbps, Tbps, and several Tbps link speeds by mmWave, THz, and indoor FSO communications, respectively. Notice that directivity and penetration losses become more significant as we move up in the spectrum chart. Although highly directed links require LoS links, they help in compensating transmission losses and mitigating the interference. Directed links and high penetration losses especially paves the way for an enhanced physical layer security. While mmWave suffers from thermal noise at the receiver, THz receivers experience both absorption and thermal noises. However, the dominant interference in FSO receivers is caused by the ambient light sources. As mmWave has already been considered as a key technology for 5G networks and beyond, the size and cost of mmWave modules are expected to reduce in short to medium period of time. Noting that commercial outdoor FSO modules are already available, tolerant DCN indoor environment further eliminates the need for bulky and costly FSO transceivers. Even though state-of-the-art THz transceivers and antennas are not very well commercialized yet, they can be foreseen to have a small form factor at reasonable costs.         

\section{Advances and Challenges in Wireless DCNs}
\label{sec:advances}
In this section, we present the challenges of implementing a real-life WDCNs along with the recent advances.
\subsection{Physical Topology Design}
\label{sec:phy}
In wired DCNs, conventional grid-based arrangement of racks is preferable to efficiently use the space. However, the need for freeing the wireless DCNs from the LoS link obstructions necessitates a paradigm shift in the physical topology (PhyTop) design, which indeed draws the upper-bound on the performance achievable by any virtual topology (VirTop). Therefore, a proper PhyTop must account for many practical concerns including scalability, reconfigurability, LoS connectivity, fault tolerance, space utilization, the small form factor of DCN components, heat and airflow, cooling cost and efficiency, and cost-performance index.

Flyways was the first to show the feasibility of applying 60-GHz technology to augment the limitations of wired DCNs by establishing on demand wireless links \cite{Flyways}. Flyway performance was further improved via indirect LoS links by reflecting signals on a ceiling mirror \cite{ceilingmirror}. Even though these proof of concept works are valuable and motivating, they are somehow topology independent and do not involve in the practical challenges.

\begin{figure*}[t]
\begin{center}
\includegraphics[width=1\textwidth]{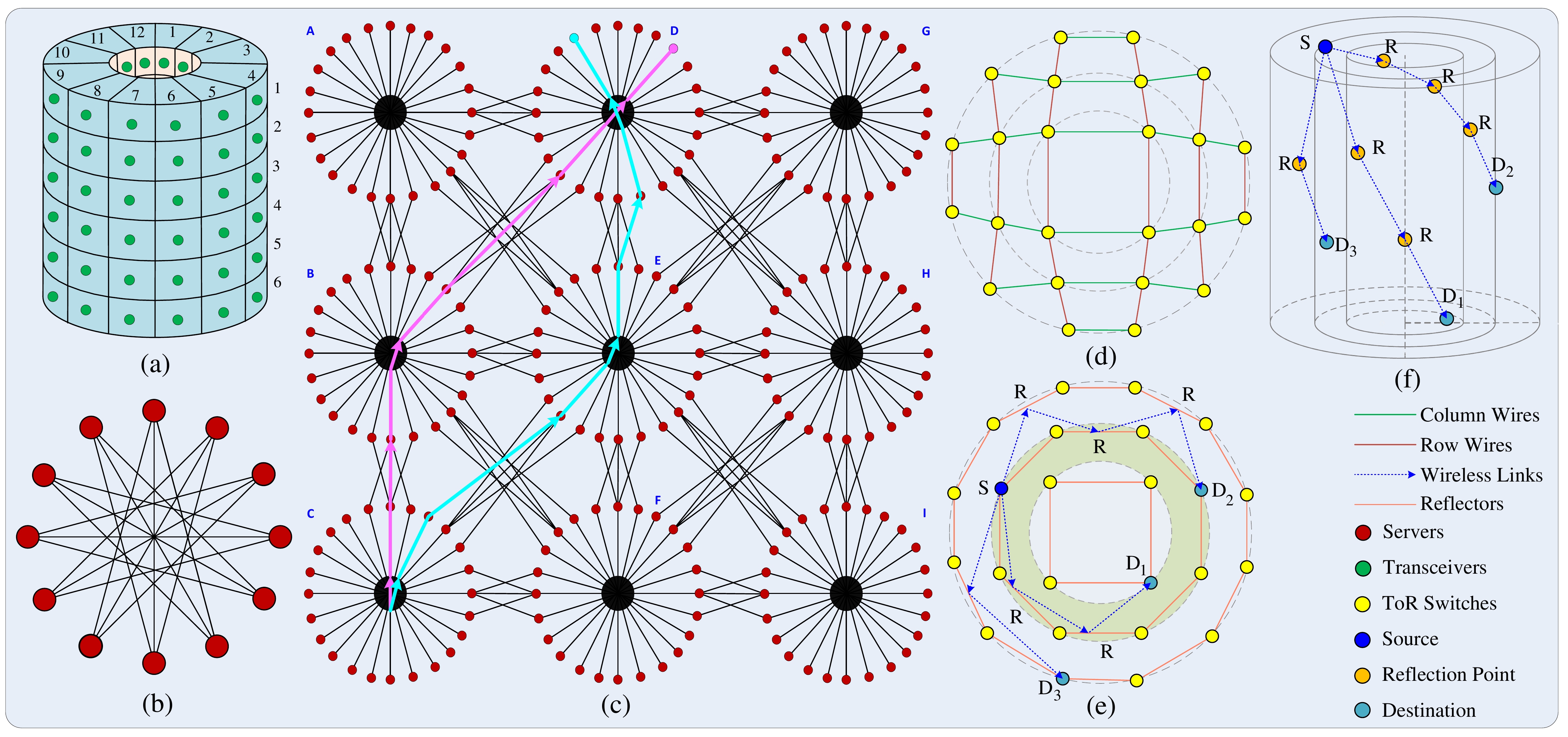}
\caption{\footnotesize Cayley~\cite{cayley} and Diamond~\cite{diamond} PhyTops for mmWave Communication: a) A cylindrical rack with 6 stories and 12 server slices, b) A 3-degree CG representing the connectivity of servers within a story, c) The bird-eye view of routing in the CDCs, d) Top-view of a bucket, e) Illustration of routing over reflectors (top-view), and  f) Side-view of a bucket and the routings in (e).}
\label{fig:mmwave_top}
\end{center}
\end{figure*}

Shin \textit{et. al.} proposed a novel concept of cylindric shaped rack design \cite{cayley} as shown in Fig. \ref{fig:mmwave_top}.a where a cylindrical rack has 6 stories and 12 slices, i.e., a total of 72 prism-shaped containers. Each container encapsulates a commodity blade server that is equipped with two 60-GHz transceivers; one for intra-rack and the other for inter-rack communications. As depicted in Fig. \ref{fig:mmwave_top}.b, intra-rack transceivers within a story forms a \textit{k}-degree \textit{Cayley Graph} (CG) where \textit{k} is a design parameter that changes with the beamwidth. Hence, entire transceivers generalize to a mesh of CG at each story [c.f. Fig. \ref{fig:mmwave_top}.c]. The Cayley DC (CDC) utilizes the space in an efficient manner, simplifies construction and maintenance, improves the bandwidth and latency, and enhances the fault tolerance by providing abundant switch-less routing paths. Authors also propose a distributed three dimensional routing protocol which calculates path traversing across servers and stories to reach the destination. If a server is identified by a tuple of rack, server, and story, e.g. $(r,s,t)$, cyan and magenta colored paths in Fig. \ref{fig:mmwave_top}.c are possible routes from $(A,12,1) \to (D,22,6)$ and $(A,12,6) \to (D,3,1)$, respectively. 
Noting that the CDC is a pure server-centric design, resulting latency and energy cost of multihop transmission exacerbates as the DCN size enlarges. Inspired by \cite{cayley}, the work in \cite{cayleyball} modifies the Cayley DCNs by using spherical racks for the sake of a higher fault tolerance of the intra-rack communications. However, it is expected to have practical ramifications in implementing such irregular shapes. 
\begin{table}
\caption{\footnotesize Comparison of Physical Topologies}
\label{tab:comp_phytop}
\centering
\resizebox{0.5\textwidth}{!}{
\begin{tabular}{cccccc}
\cline{2-6}
\multicolumn{1}{l|}{}   & \multicolumn{5}{c|}{\cellcolor[HTML]{68CBD0}\textbf{Physical Topology}}  \\ \hline
\rowcolor[HTML]{00D2CB} 
\multicolumn{1}{|c|}{\cellcolor[HTML]{9698ED}\textbf{Attributes}} & \multicolumn{1}{c|}{\cellcolor[HTML]{00D2CB}Row}          & \multicolumn{1}{c|}{\cellcolor[HTML]{00D2CB}FlyX} & \multicolumn{1}{c|}{\cellcolor[HTML]{00D2CB}Cayley}       & \multicolumn{1}{c|}{\cellcolor[HTML]{00D2CB}Diamond}      & \multicolumn{1}{c|}{\cellcolor[HTML]{00D2CB}OWCell}       \\ \hline \vspace{-5pt}
\cellcolor[HTML]{FFFFFF}                                          & \cellcolor[HTML]{FFFFFF}                                  & \cellcolor[HTML]{FFFFFF}                                              & \cellcolor[HTML]{FFFFFF}                                  &                                                           &                                                           \\ \hline
\rowcolor[HTML]{ECF4FF} 
\multicolumn{1}{|c|}{\cellcolor[HTML]{CBCEFB}Scalability}         & \multicolumn{1}{c|}{\cellcolor[HTML]{ECF4FF}$\checkmark$} & \multicolumn{1}{c|}{\cellcolor[HTML]{ECF4FF}$\times$}                 & \multicolumn{1}{c|}{\cellcolor[HTML]{ECF4FF}$\checkmark$} & \multicolumn{1}{c|}{\cellcolor[HTML]{ECF4FF}$\checkmark$} & \multicolumn{1}{c|}{\cellcolor[HTML]{ECF4FF}$\checkmark$} \\ \hline
\rowcolor[HTML]{EFEFEF} 
\multicolumn{1}{|c|}{\cellcolor[HTML]{CBCEFB}Reconfigurability}   & \multicolumn{1}{c|}{\cellcolor[HTML]{EFEFEF}$\times$}     & \multicolumn{1}{c|}{\cellcolor[HTML]{EFEFEF}$\times$}                 & \multicolumn{1}{c|}{\cellcolor[HTML]{EFEFEF}$\checkmark$} & \multicolumn{1}{c|}{\cellcolor[HTML]{EFEFEF}$\checkmark$} & \multicolumn{1}{c|}{\cellcolor[HTML]{EFEFEF}$\checkmark$} \\ \hline
\rowcolor[HTML]{ECF4FF} 
\multicolumn{1}{|c|}{\cellcolor[HTML]{CBCEFB}LoS Connectivity}    & \multicolumn{1}{c|}{\cellcolor[HTML]{ECF4FF}Low}     & \multicolumn{1}{c|}{\cellcolor[HTML]{ECF4FF}Medium}                 & \multicolumn{1}{c|}{\cellcolor[HTML]{ECF4FF}High} & \multicolumn{1}{c|}{\cellcolor[HTML]{ECF4FF}High} & \multicolumn{1}{c|}{\cellcolor[HTML]{ECF4FF} High} \\ \hline
\rowcolor[HTML]{EFEFEF} 
\multicolumn{1}{|c|}{\cellcolor[HTML]{CBCEFB}Interference}        & \multicolumn{1}{c|}{\cellcolor[HTML]{EFEFEF}High}         & \multicolumn{1}{c|}{\cellcolor[HTML]{EFEFEF}High}                     & \multicolumn{1}{c|}{\cellcolor[HTML]{EFEFEF}Low}          & \multicolumn{1}{c|}{\cellcolor[HTML]{EFEFEF}Low}          & \multicolumn{1}{c|}{\cellcolor[HTML]{EFEFEF}Low}          \\ \hline
\rowcolor[HTML]{ECF4FF} 
\multicolumn{1}{|c|}{\cellcolor[HTML]{CBCEFB}Space Utilization}   & \multicolumn{1}{c|}{\cellcolor[HTML]{ECF4FF}High}         & \multicolumn{1}{c|}{\cellcolor[HTML]{ECF4FF}Low}                      & \multicolumn{1}{c|}{\cellcolor[HTML]{ECF4FF}High}         & \multicolumn{1}{c|}{\cellcolor[HTML]{ECF4FF}High}         & \multicolumn{1}{c|}{\cellcolor[HTML]{ECF4FF}Medium}       \\ \hline
\rowcolor[HTML]{ECF4FF} 
\multicolumn{1}{|c|}{\cellcolor[HTML]{CBCEFB}Bandwidth}           & \multicolumn{1}{c|}{\cellcolor[HTML]{EFEFEF}Low}          & \multicolumn{1}{c|}{\cellcolor[HTML]{EFEFEF}Low}                      & \multicolumn{1}{c|}{\cellcolor[HTML]{EFEFEF}High}         & \multicolumn{1}{c|}{\cellcolor[HTML]{EFEFEF}High}         & \multicolumn{1}{c|}{\cellcolor[HTML]{EFEFEF}High}         \\ \hline
\rowcolor[HTML]{EFEFEF} 
\multicolumn{1}{|c|}{\cellcolor[HTML]{CBCEFB}Small Form Factor}   & \multicolumn{1}{c|}{\cellcolor[HTML]{ECF4FF}$\checkmark$} & \multicolumn{1}{c|}{\cellcolor[HTML]{ECF4FF}$\times$}                 & \multicolumn{1}{c|}{\cellcolor[HTML]{ECF4FF}$\checkmark$} & \multicolumn{1}{c|}{\cellcolor[HTML]{ECF4FF}$\checkmark$} & \multicolumn{1}{c|}{\cellcolor[HTML]{ECF4FF}$\checkmark$} \\ \hline
\rowcolor[HTML]{EFEFEF} 
\multicolumn{1}{|c|}{\cellcolor[HTML]{CBCEFB}Need for Switch}     & \multicolumn{1}{c|}{\cellcolor[HTML]{EFEFEF}$\checkmark$} & \multicolumn{1}{c|}{\cellcolor[HTML]{EFEFEF}$\checkmark$}             & \multicolumn{1}{c|}{\cellcolor[HTML]{EFEFEF}$\times$}     & \multicolumn{1}{c|}{\cellcolor[HTML]{EFEFEF}$\checkmark$} & \multicolumn{1}{c|}{\cellcolor[HTML]{EFEFEF}$\checkmark$} \\ \hline
\multicolumn{1}{l}{}                                              & \multicolumn{1}{l}{}                                      & \multicolumn{1}{l}{}                                                  & \multicolumn{1}{l}{}                                      & \multicolumn{1}{l}{}                                      & \multicolumn{1}{l}{}                                      \\
\multicolumn{1}{l}{}                                              & \multicolumn{1}{l}{}                                      & \multicolumn{1}{l}{}                                                  & \multicolumn{1}{l}{}                                      & \multicolumn{1}{l}{}                                      & \multicolumn{1}{l}{}                                     
\end{tabular}
}
\end{table}

Diamond DC (DDC) is another interesting PhyTop design which hybridizes the wired and 60-GHz wireless DCNs \cite{diamond}. As demonstrated in Fig. \ref{fig:mmwave_top}.d, DDC arranges and wires the racks based on a herringbone pattern which has $4i$ edges at the $i^{th}$ ring, which is referred to as buckets. Fig.~\ref{fig:mmwave_top}.d and Fig.~\ref{fig:mmwave_top}.f illustrate top and side views of a bucket with three rings, respectively. To prevent blocking and interference, the gap between the racks is used as a ring reflection space (RRS) by equipping the edges with the reflectors at each layer (similar to stories in \cite{cayley}). In this way, a rack in the ring $R_i$ can reach to other racks within rings $R_{i-1}$ and $R_{i+1}$ by making the multiple reflections in the RSS whereas the remaining connections are possible via the herringbone wires. The side and top views of possible routes between a source server and destination server $D_1$/$D_2$/$D_3$ in the ring $R_1$/$R_2$/$R_3$ are illuminated in Fig. \ref{fig:mmwave_top}.e and Fig. \ref{fig:mmwave_top}.f, respectively. The DDC offers a good scalability as the buckets can be arranged to form a grid similar to the CDC grids in Fig. \ref{fig:mmwave_top}.c. Hybridization of wired and wireless connections increases the fault tolerance and provides ample routing opportunities. Both CDCs and DDCs are inspiring examples of achieving a blockage-free DCN while utilizing the space in an efficient manner.  

\begin{figure*}[t]
\begin{center}
\includegraphics[width=1\textwidth]{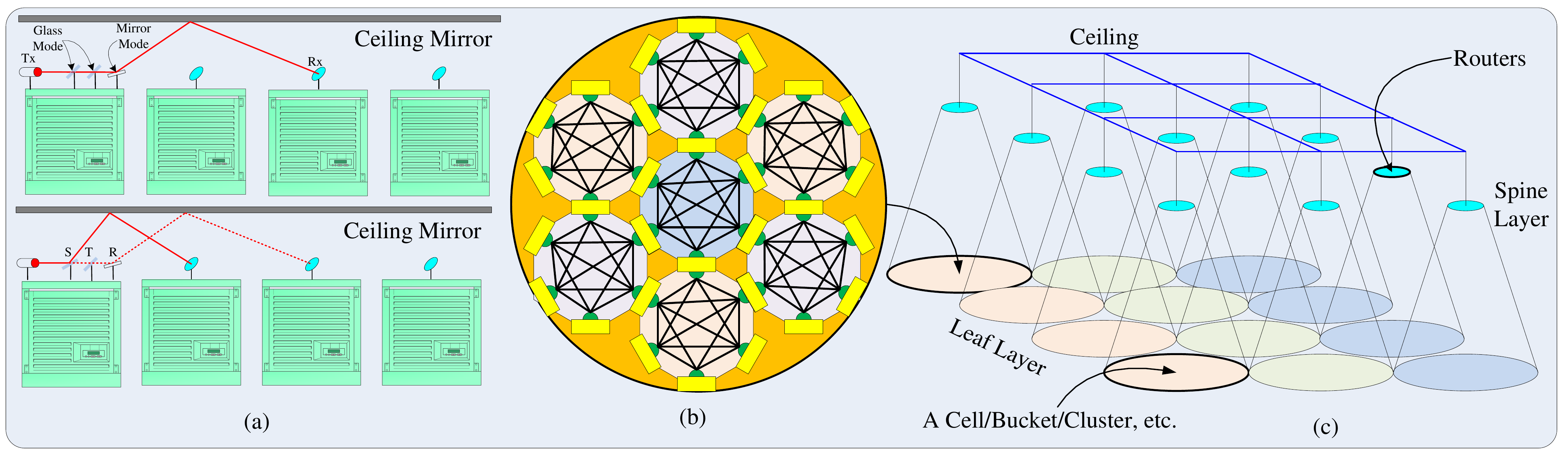}
\caption{\footnotesize Different PhyTops for OWCs: a) FireFly and FlyCast, b) OWcell, and c) Hybrid server-switch centric approach.}
\label{fig:owc_top}
\end{center}
\end{figure*}

Similar to \cite{ceilingmirror}, the first attempts to adopt OWC for DCNs considered a ceiling mirror to establish links between the ESs \cite{firefly, FlyCast}. In FireFly \cite{firefly}, authors employed pre-configured and aligned mirrors that can switch between glass and mirror modes. Using three different modes (i.e., reflective, transmissive, and splitting modes), FlyCast extends the FireFly to enable multicast communication without using a switch. Based on the grid-based PhyTop, authors of \cite{VLC} interconnect the rack grid using stationary VLC receivers. They also consider randomly placed VLC receivers atop the racks for adapting the topology to accommodate the traffic. However, the limited ceiling height and the clearance above the racks hinders the application of these techniques in practice. Moreover, typical rack size delimits the total number of OWC transceivers that can be placed atop the racks, which may also not be desirable due to the increasing interference. Alternatively, OWCell uses the cell of racks as a building block to create mega DCNs \cite{OWCell}. In order to facilitate LoS links, racks within a cell are arranged in regular polygonal shapes and fully connected to each other. Fig.~\ref{fig:owc_top}.b shows an OWCell topology consisting of 7 hexagonal shaped cells. Although OWCell has desirable scalability and connectivity, its space utilization is not as efficient as the CDCs and DDCs. For reader's convenience, we tabulate the comparison of PhyTops in Table~\ref{tab:comp_phytop}.

\subsection{Virtual Topology Design}
\label{sec:virt} 

VirTop design is essential for leveraging the reconfigurability and flexibility of wireless DCNs to meet the dynamically changing QoS demands of network entities. In this respect, VirTop design involves many interesting and challenging open research areas such as resource allocation, scheduling, power control, beamforming, interference management, routing, etc.  Even though early research efforts put a more focus on PhyTop design issues, there exists many open research problems related to VirTop design:

First and foremost, directed nature of the candidate wireless technologies necessitates a paradigm shift in the definition of a resource block (RB) to include space as the third dimension in addition to time and frequency. Taking the large operational bandwidths into account, it is important to standardize frequency and time slots in a proper granularity in order to have a sufficiently large number of concurrent transmissions. When frequency-time blocks are spatially reused by the narrow-beam transmissions, resulted three-dimensional (3D) RBs would significantly improve the network performance. Although using RBs in an orthogonal fashion can ease the resource allocation problem, it contradicts with the high area spectral efficiency of the 3D-RBs. Instead, it is appealing to develop resource allocation strategies which reuse the frequency by managing the interference. At this point, beamforming and power control mechanisms are required to ensure that concurrent transmissions on the same frequency-time blocks do not violate tolerable interference limits of one another. 

In order to orchestrate all these network functions, a 3D scheduling is imperative to reap the full benefits of the reconfigurability and flexibility of WDCNs. Scheduling algorithms are required to be adaptive to account for the dynamically changing traffic patterns and QoS requirements of different flow classes. Furthermore, scheduling should be handled in coordination with the routing protocols, especially in the heavy traffic loads. That is, scheduling mechanisms should exploit the available network resources to such an extent that there exist available paths to keep servers and racks are connected. 

Scheduling can be implemented either locally or globally. In global scheduling, network entities are controlled by a central unit which is aware of the entire network state, which makes joint contention/congestion control and routing possible for a desirable end-to-end performance. The control message size, message passing frequency, and distance increase with the size of the DCN, which eventually yields higher time, space, and communication overhead, respectively. On the other hand, local scheduling can deal with the congestion given the neighborhood information. Also noting that the limited performance of distributed solutions, a hybrid approach would be more practical. For instance, the distributed methods could be sufficient to serve the need of routine traffic patterns and to handle the self-interference of inter-rack communications by spatially reusing the RBs in a dynamic manner. On top of this, a centralized solution can create on demand paths to leverage the global network information to mitigate the chronic problems of oversubscription and hotspots. 

\subsection{QoS Provisioning and Flow Classification} 
\label{sec:FC}
QoS provisioning is a critical component that must be guaranteed for highly diversified types of DCN traffic. For instance, a regular cell phone call and blue light box call cannot be treated with the same priority and QoS constraints. Therefore, it is important to distinguish flows based on size, completion time requests, priority, etc. DC flows are typically classified as bandwidth-hungry elephant flows (EFs) or delay-sensitive mice flows (MFs). While the majority of the traffic volume is carried out by EFs, the majority of flows are typically MFs. Among the MFs, there might be higher priority critical flows with ultra-reliable and low-latency service demands. Thus, VirTop should be designed to distinguish and treat different types of flows based on their needs. For instance, MFs may experience intolerable delays if they are routed along with EFs on the same path.  Hence, flow detection mechanism should be rapid, precise, and low-overhead to prevent such performance degradations, which cannot be satisfied by the existing techniques (e.g., packet sampling, port mirroring, etc.).


Accordingly, we developed a distributed fast, lightweight, and accurate flow detection detection mechanism, namely LightFD \cite{LightFD}. By leveraging the TCP behavior, LightFD is designed as a module installed in the virtual-switches. However, distributed solutions are only suitable for modifiable switches with location privileges. In this paper, we then propose an alternative centralized scheme where a central unit reconfigures the virtual-switches by remotely communicating with switches via OpenFlow protocol. Based on extensive emulation results, Fig. \ref{fig:tg} shows that the accuracy of LightFD schemes outperforms the traditional methods. Moreover, the detection speed of the distributed and centralized schemes are in the order of milliseconds and ten milliseconds, respectively, which is indeed promising in comparison with the OpenSample. The performance degradation of centralized scheme is due to the communication and computation overhead. 


\subsection{Data Grooming (DG) and Load Balancing (LB)}
\label{sec:TG}
The complexity of VirTop design increases not only with the size but also with the massive and diverse amount of traffic generated across the DCN. By bundling and treating the same class of flows destined to a common destination, DG can increase the bandwidth utilization and relax the complexity of the network management. In this direction, we develop a three-step DG (3SDG) policy for a two-layer WDCN topology in \cite{tg} where leaf and spine layers respectively comprise of ESs and CSs, which are interconnected via WDM-FSO links [cf. Fig. \ref{fig:owc_top}].

In 3SDG, servers first groom MFs destined to the same server. Then, servers re-grooms the flows in the first step intended to the same rack. Finally, ESs performs a final grooming to create rack-to-rack (R2R) jumbo flows. Since MFs forms the majority of the generated flows, 3SDG calculates predetermined R2R routing paths whose capacity is calculated based on the flow arrival statistics, flow size, and required flow completion duration. In this manner, a significant portion of flows is handled with a very low complexity by ensuring the QoS demands. The remaining wavelengths (i.e., channels or RBs) are exploited by EFs which are transferred over server-to-server (S2S) express routes. By routing MFs and EFs on disjoint paths, 3SDG prevents highly probable congestion occasions as delay sensitive MFs can be affected by bandwidth-hungry EFs easily. On top of this, 3SDG immediately serves mission critical flows over available S2S routes by interrupting the ongoing EFs. Noting the small size of the MFs and available large capacity, such an impulsive interruption does not harm the EF performance distinguishably.

To comply with the DCN traffic dynamics, changing the VirTop frequently may not yield a desirable performance because of delays caused by calculations, hardware reconfigurations, and waiting for flow completions. Alternatively, LB techniques may be regarded as a soft reconfiguration tool by migrating congested flows to underutilized paths. Combining hard and soft reconfigurations can shrink delays and boost the overall bandwidth utilization via carefully designed LB schemes. Therefore, we extend our work in \cite{tg} by considering an extra flow class, cat flows (CFs), whose size ranges between that of MFs and EFs. To balance the load without changing hardware parameters, the proposed LB scheme migrates CFs and EFs within and across two disjoint VirTops, each dedicated to MFs and EFs.  
\begin{figure}
    \centering
        \includegraphics[width=0.5\textwidth]{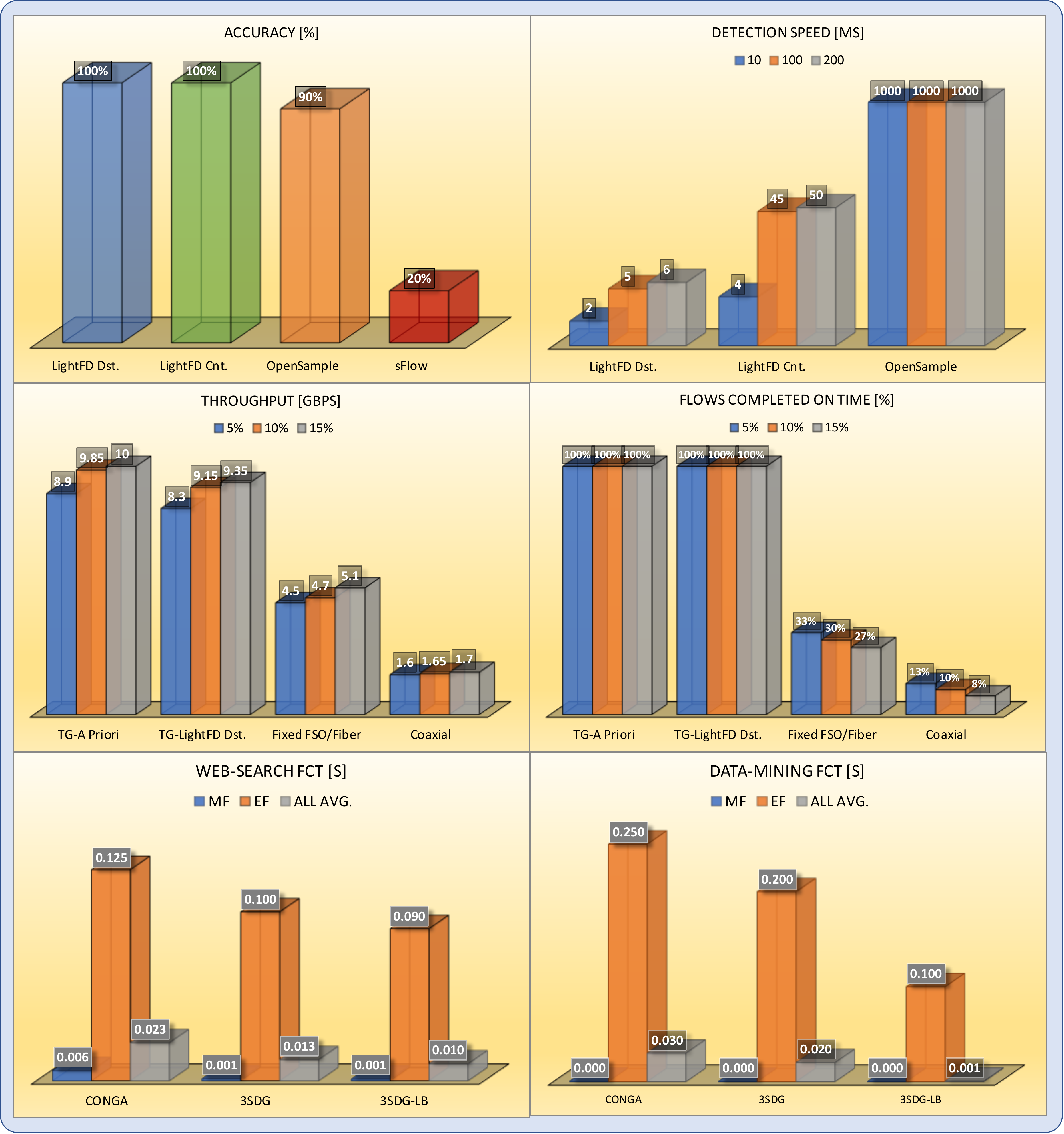}
    \caption{Performance of LightFD and the proposed TG policy.}\label{fig:tg}
\end{figure}

In Fig. \ref{fig:tg}, throughput of MFs and EFs are illustrated for four different cases: 1) \textit{TG-A priori}: 3SDG with given flow classifications; 2) \textit{TG-LightFD Dst.}: 3SDG with the distributed LightFD; 3) \textit{Fixed FSO/Fiber}: Fiber links or FSO wavelengths with fixed capacity; and 4) \textit{Coaxial}:  the traditional wired DCNs. Noting that case-3 and case-4 exploits equal cost multipath routing, the capacity of coaxial cables and each wavelengths are set to 1 Gbps and 10 Gbps, respectively. Apparently, flexibility and reconfigurability of the WDCNs provide a superior performance in comparison with the wired/fixed counterparts. Moreover, the impact of overhead introduced by the flow detection mechanism can be distinguished easily. Comparing the detection speed and accuracy of other schemes, the performance loss of TG-LightFD would be more significant if other detection schemes were employed. For web-search and data-mining workloads, Fig. \ref{fig:tg} also depicts that the percentage of on-time flows decreases significantly in wired DCNs due to the low utilization result of inflexibility. Moreover, Fig. \ref{fig:tg} compares the flow completion time (FCT) of 3DSG with the proposed LB scheme (3DSG-LB) and CONGA \cite{CONGA}. Results show that 3DSG outperforms CONGA thanks to its inherent LB capability. 3DSG performance is further enhanced by 3DSG-LB which is more obvious in FCT of EFs. However, the difference between 3DSG and 3DSG-LB for MFs since the 3DSG is already able to handle MF loads by allocating dedicated paths with required capacity.  

\section{Conclusions and Future Prospects}
\label{sec:prospects}

In this article, we provide an overview of WDCNs for embracing the emerging information and communication technologies that require storage and processing of the highly diverse massive amount of data. After briefing and comparing the virtues and drawbacks of potential high-speed wireless technologies, we have presented advances and challenges of implementing WDCNs from both theoretical and practical points of views. Following the presentation of ongoing research efforts focused on PhyTop design, we also discussed overlooked aspects such as VirTop design, flow detection, data grooming, and load balancing. Some of the promising directions and open problems for future research are outlined as follows:

\paragraph{Hybridization of PhyTops \& Wireless Technologies}   
So far, proposed PhyTops either tries to minimize the use of switches or eliminate them. However, a large number of hops may be required in mega DCNs to reach a distant target, which exploits the energy, bandwidth, and processing power of the network entities along the path. Alternatively, assigning router or access points to each group of racks (i.e., cell \cite{OWCell}, bucket \cite{diamond}, cluster \cite{cayley}) would yield a better performance and scalability [cf. Fig. \ref{fig:owc_top}. c]. These routers can be interconnected either with fibers or WDM-FSO links to provide a backbone to carry the aggregate traffic. Even though proposed PhyTops are merely presented for a certain wireless technology, indeed their directed nature paves the way for a combination of these techniques. For instance, the VLC or THz communication would be a perfect fit for both intra-rack and nearby inter-rack connections. A good option to connect ToR switches with routers may be using FSO links. In this case, these out of band technologies would not interfere with each other. Our overall message is that there are plentiful interesting research problems to be extracted from the hybridization of wireless technologies and underlying PhyTops.  

\paragraph{In-depth Exploration of VirTop Aspects}
VirTop design can be cast as a joint optimization of subproblems mentioned in Section \ref{sec:advances}. Unfortunately, solving this joint problem requires significant computational power and time complexity, which increase with the scale of DCs. Although research community has shed light on the PhyTop issues to prove the concept of WDCN as a first step, the real benefits of a promising PhyTop can only be provided by a proper VirTop design, which is not explored in depth yet. In particular, given the real historical DCN traces, probabilistic tools, machine learning, and stochastic optimization can be employed to deal with the dynamic nature of the DCN traffic. 


\paragraph{Inter-DCN Wireless Communications}
Companies establish geographically distributed DCs for many reasons such as content disaster recovery, load balancing, making data available to the end users as close as possible, etc. Data or application migration among these DCs requires ultra high-speed and low-latency connections. Although wireless communications may not be a strong alternative to fibers for connecting DCNs across a continent, it is a promising solution to use on-demand or constant wireless links to form a cluster of DCNs.
This is especially useful in the metropolitan areas where the cost and complexity associated with laying fiber cables is high. In addition to intra-DCN wireless communications, investigating the integration of pure or hybrid wireless intra-DCN and inter-DCN is a potential future research direction with many open interesting problems.   

\paragraph{Underwater Containerized DCNs}
Containerized DCNs have recently attracted attention thanks to their mobility, modularity, and cost-efficiency. Although wiring cost and complexity may not be a source of concern in containerized DCNs, flexibility and reconfigurability can still be the main motivation for using wireless communications. Given the large bandwidth availability of candidate wireless solutions and the short distance within a container, interference-free high-speed wireless links can be established and managed with low overhead. In June 2018, the launch of project Natick of Microsoft (\url{https://natick.research.microsoft.com}) revolutionized the use of containerized DCNs by sinking cylindrical tube shaped DCNs into the sea for the sake of green power offered by offshore renewable energy sources, significantly reduced cooling costs, and gratifying proximity to where half of the globe’s population inhabits. Underwater OWC is an emerging field of research as a substitute for acoustic systems with very high data rates and low-latency \cite{uownsurvey}. Even if containerized DCNs can be interconnected via fibers, on-demand underwater OWC links can provide the required flexibility and reconfigurability for desirable performance. 

\paragraph{Security and Privacy}
Nowadays it is a common practice for companies to rent a portion of mega DCNs in order to prevent maintaining and upgrading costs of a self-owned enterprise network. Although some customers need services in the application layers, some others may require entire control of the rented partition of the DCN. In this case, it is critical to isolate wirelessly transmitted data from unintended nodes in order to prevent security and privacy risks. Fortunately, the directivity and limited penetration of high-speed wireless links make it possible to partition the DCN via physical obstacle, for example, a rented OWCell can be put in a cylindrical container and isolated from the rest of the DCN. Such an approach may enhance the immunity to eavesdropping, allow to develop low-overhead security protocols, and hence increase overall networking performance. Security and privacy are in the center of skepticism of the cloud service providers toward the WDCNs as they are legally responsible for protecting the sensitive information.

\paragraph{Analysis of Cost-Performance Tradeoff}
Cost-performance index is a critical factor to be taken into account in the design and provisioning of WDCNs. From the monetary cost point of view, it is necessary to compare the hardware cost of cabling and wireless modules to decide on the suitable wireless technology and underlying PhyTop. Additionally, DCNs are extremely power hungry infrastructures with a considerable carbon-footprint, which dictates energy-efficiency as an important design parameter. Since switches are the most expensive and power consuming units, there is an urgent need to quantify the power reductions offered by WDCNs via elimination of switches. While cables are passive components with a power loss in a certain distance, their indirect power cost could be significant because of the additional power spent to compensate heat dissipation and extra space required for large cable bundles. On the other hand, wireless modules actively consume energy for both transmission and digital signal processing. Hence, VirTop should account for the energy cost while reconfiguring the DCN for changing demands.   

\paragraph{Controller Placement Problem}
In software-defined network driven DCNs, decoupling the control and data planes yields an interesting problem: deciding on the number of controllers and their locations. This problem is closely related with the reliability, scalability, and latency of the control plane signals. Indeed, this can also be alleviated by employing wireless links. Instead of using highly directed wireless technologies, microwave communications would be more suitable to control signaling for two reasons: 1) Control signals propagating in the microwave band has no interference on out-of-band data plane and 2) Microwave communications can provide ubiquitous connection to all racks thanks to its omnidirectional nature, which is necessary for broadcasting commands to all network entities. However, integration of wireless control and data planes are still not investigated yet.  

\bibliographystyle{IEEEtran.bst}
\bibliography{twocolumn}
\newpage
\begin{IEEEbiography}{Abdulkadir Celik}(S'14-M'16) received the B.S. degree in electrical-electronics engineering from Selcuk University in 2009, the M.S. degree in electrical engineering in 2013, the M.S. degree in computer engineering in 2015, and the Ph.D. degree in co-majors of electrical engineering and computer engineering in 2016, all from Iowa State University, Ames, IA. He is currently a postdoctoral research fellow at Communication Theory Laboratory of King Abdullah University of Science and Technology (KAUST). His current research interests include but not limited to 5G and beyond, wireless data centers, UAV assisted cellular and IoT networks, and underwater optical wireless communications, networking, and localization. 

\end{IEEEbiography}
\vspace*{-2\baselineskip}
\begin{IEEEbiography}{Basem Shihada}(SM'12) is an Associate and Founding Professor of computer science and electrical engineering in the Computer, Electrical and Mathematical Sciences \& Engineering (CEMSE) Division at King Abdullah University of Science and Technology (KAUST). Before joining KAUST in 2009, he was a visiting faculty at the Computer Science Department in Stanford University. His current research covers a range of topics in energy and resource allocation in wired and wireless communication networks, including wireless mesh, wireless sensor, multimedia, and optical networks. He is also interested in SDNs, IoT, and cloud computing. 
\end{IEEEbiography}
\vspace*{-2\baselineskip}
\begin{IEEEbiography}{Mohamed-Slim Alouini} (S'94-M'98-SM'03-F'09) was born in Tunis, Tunisia. He received the Ph.D. degree in Electrical Engineering from the California Institute of Technology (Caltech), Pasadena, CA, USA, in 1998. He served as a faculty member in the University of Minnesota, Minneapolis, MN, USA, then in the Texas A\&M University at Qatar, Education City, Doha, Qatar before joining King Abdullah University of Science and Technology (KAUST), Thuwal, Makkah Province, Saudi Arabia as a Professor of Electrical Engineering in 2009. His current research interests include the modeling, design, and performance analysis of wireless communication systems.
\end{IEEEbiography}

\end{document}